\documentstyle[epsf]{aipproc}

\catcode`@=11
\long\def\@makefntext#1{\parindent 0pt\hsize\columnwidth\parskip0pt\relax
\footnotesize\baselineskip12pt\def\strut{\vrule width0pt height0pt depth1.75pt\relax}%
\mbox{$\m@th^{\@thefnmark}$\hspace*{3pt}}#1}
\catcode`@=12

\begin{document}

\font\fortssbx=cmssbx10 scaled \magstep2
\hbox to \hsize{{\fortssbx University of Wisconsin - Madison}
\hfill
\vtop{\hbox{\bf MADPH-00-1165}
      \hbox{February 2000}}}

\title{Neutrino Masses and Mixing\\ at the Millennium\footnote{Talk presented at the {\it 5th International Conference on Physics Potential \& Development of $\mu^+\mu^-$ Colliders}, San Francisco, December 1999.}}

\author{V. Barger}

\address{\vglue-3ex
Physics Department, University of Wisconsin, Madison, WI 53706\vglue-3ex}

\maketitle

\begin{abstract}
Recent evidence for neutrino oscillations has revolutionized the study of neutrino masses and mixing. This report gives an overview of what we are learning from the neutrino oscillation experiments, the prospects for the near term, and the bright future of neutrino mass studies.
\end{abstract}

\section{Neutrino Counting}
How many neutrinos are there? Neutrino counting at LEP of $Z\to\nu\bar\nu$ decays obtains $N_\nu=3$ active flavors --- the expected $\nu_e,\ \nu_\mu,\ \nu_\tau$. However, light isosinglet, right-handed ``sterile" neutrinos with no gauge boson interactions could also exist. Big Bang Nucleosynthesis determines the equivalent number of massless neutrinos at the time of nucleosynthesis. The bounds inferred ($N_\nu < 3.2$\cite{burles}, $N_\nu < 4$\cite{lisi}, $N_\nu < 5.3$\cite{olive}) depend on which measurements of the $^4$He and D/H abundances are used in the analysis. Neutrino oscillation phenomenology for $N_\nu=3$ and $N_\nu=4$ is very different and both options need to be considered at present.

\section{Neutrino Masses}

\begin{figure}
\centering\leavevmode
\epsfxsize=3.1in\epsffile{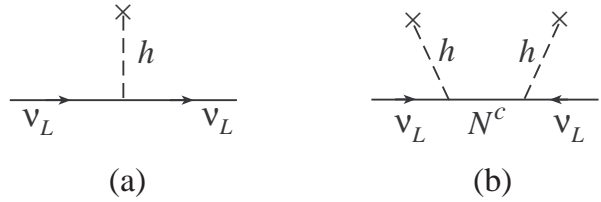}

\caption[]{\label{fig:seesaw}
Tree-level diagrams for (a)~Dirac mass and (b)~Majorana mass generation.}
\end{figure}

Tree-level mass generation occurs through the Higgs mechanism. The Dirac mass $m_D$ arises  in a lepton conserving ($\Delta L = 0$) interaction (see Fig.~\ref{fig:seesaw}) and requires a right-handed neutrino. A Majorana mass $m_M$ occurs through a $\Delta L = 2$ process with only a left-handed light neutrino field and a heavy isosinglet intermediate field $N^c$. Then the see-saw mechanism\cite{seesaw} with $m_D \sim 10^2$~GeV and $m_M > 10^{12}$~GeV generates light neutrinos
\begin{equation}
m_\nu = {m_D^2\over m_M}
\end{equation}
that are nearly Majorana.
In the case that $m_D\sim m_M\sim\rm eV$, as can be realized in some models\cite{langacker,BLLP}, active--sterile neutrino oscillations can take place.

\begin{figure}[h]
\centering\leavevmode
\epsfxsize=1.8in\epsffile{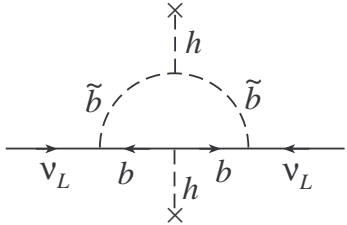}

\caption[]{\label{fig:radiative}
Radiative diagram for Majorana neutrino mass generation via $R$-parity violating interactions.}
\end{figure}

Neutrino mass can alternatively be generated radiatively by new interactions\cite{RMreview}, such as the $R$-parity violating $\nu b\tilde b$ interaction\cite{drees}, as illustrated in Fig.~\ref{fig:radiative}.

\section{Three-Neutrino Oscillatons}

The relation of the three-neutrino flavor eigenstates to the mass eigenstates is
\begin{equation}
\left(\begin{array}{c} \nu_e\\ \nu_\mu\\ \mu_\tau\end{array}\right) = 
U \left(\begin{array}{c} \nu_1\\ \nu_2\\ \nu_3\end{array}\right)\,,
\end{equation}
where $U$ is the $3\times3$ Maki-Nakagawa-Sakata (MNS) mixing matrix\cite{MNS}. It can be parametrized by
\begin{eqnarray}
U &=& \left(\begin{array}{ccc}
  c_{13} c_{12}       & c_{13} s_{12}  & s_{13} e^{-i\delta} \\
- c_{23} s_{12} - s_{13} s_{23} c_{12} e^{i\delta}
& c_{23} c_{12} - s_{13} s_{23} s_{12} e^{i\delta}
& c_{13} s_{23} \\
    s_{23} s_{12} - s_{13} c_{23} c_{12} e^{i\delta}
& - s_{23} c_{12} - s_{13} c_{23} s_{12} e^{i\delta}
& c_{13} c_{23}
\end{array}\right)\nonumber\\
  && \hspace{7em} \times
\left(\begin{array}{ccc}
1& 0& 0\\
0& e^{i\phi_2}& 0\\
0& 0& e^{i(\phi_3+\delta)}
\end{array}\right)
\end{eqnarray}
where $c_{ij}=\cos\theta_{ij}$ and $s_{ij} = \sin\theta_{ij}$. The extra diagonal phases are present for Majorana neutrinos but do not affect oscillation phenomena.

With three neutrinos there are two independent $\delta m^2$ and $\delta m_a^2 \gg \delta m_b^2$ is indicated by the oscillation evidence. The vacuum oscillation probabilities are
\begin{equation}
P(\nu_\alpha \to \nu_\beta) = A_{\alpha\beta} \sin^2 \Delta_a - B_{\alpha\beta} 
\sin^2\Delta_b + \epsilon_{\alpha\beta} J \sin2\Delta_b \,,
\end{equation}
where $\Delta_a \equiv \delta m_a^2 L / 4E_\nu$. $A_{\alpha\beta}$ is the amplitude of the leading oscillation, $B_{\alpha\beta}$ the amplitude of the sub-leading oscillation and $J$ the CP-violating amplitude; all are determined by the $U$ matrix elements. The physical variable is $L/E_\nu$, where $L$ is the baseline from source to detector and $E_\nu$ is the neutrino energy.

\section{Matter Effects}

In matter, $\nu_e$ scatter differently from $\nu_\mu$ and $\nu_\tau$\cite{wolf}, and the effective neutrino mixing amplitude in matter can be very different from the vacuum amplitude\cite{BPPW,MS}. For the leading oscillation  the matter and vacuum oscillation mixings are related in the approximation of constant matter density by
\begin{equation}
\sin^22\theta_{13}^m = {\sin^22\theta_{13}\over 
\left(\cos2\theta_{13} - A/\delta m_a^2\right)^2 + \sin^22\theta_{13}}\,,
\end{equation}
where
\begin{equation}
A = 2\sqrt2\, G_F \, Y_e\, \rho(x)\,E_\nu \,.
\end{equation}
Here $Y_e$ is the electron fraction and $\rho(x)$ is the density at path-length $x$. The $\nu_e\to\nu_\mu$ (or $\nu_e\to\nu_\tau$) oscillation argument in matter of constant density is
\begin{equation}
\Delta_a^m = {1.27\delta m_a^2({\rm eV^2})\, L({\rm km})\over E_\nu(\rm GeV)} \; \sqrt{\left({A\over\delta m_a^2} - \cos2\theta_{13}\right)^2 + \sin^22\theta_{13}}\,.
\end{equation}

Resonance enhancements of the oscillation amplitude are possible for $\delta m_a^2 > 0$, while suppression occurs for $\delta m_a^2 < 0$. It is significant that the resonant energies correspond to neutrino energies relevant to the atmospheric and solar anomalies.
\begin{eqnarray}
&{\rm Earth:}&\quad E_\nu\simeq 15{\rm~GeV} \left(\delta m_a^2\over 3.5\times10^{-3}{\rm\,eV^2}\right) \left(1.5{\rm~gm/cm^3}\over \rho Y_e\right)
\\
&\rm Sun:&\quad E_\nu\simeq 10{\rm\ MeV} \left( \delta m_b^2\over 10^{-5}{\rm \,eV^2} \right) \left( 10{\rm~g/cm^3}\over \rho Y_e \right)\,.
\end{eqnarray}

\section{Atmospheric Neutrino Oscillations}

\begin{figure}[t]
\centering\leavevmode
\epsfxsize=3.6in\epsffile{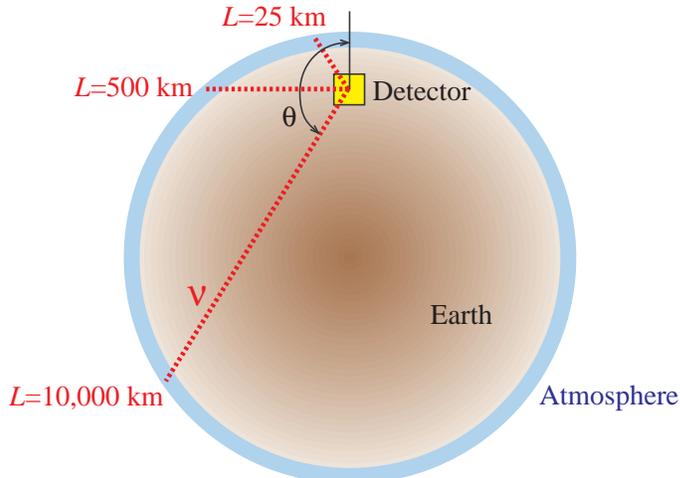}

\bigskip
\caption[]{\label{fig:depend}
Baseline dependence on the zenith angle in atmospheric neutrino experiments.}
\end{figure}

\begin{figure}[t]
\centering\leavevmode
\epsfxsize=2in\epsffile{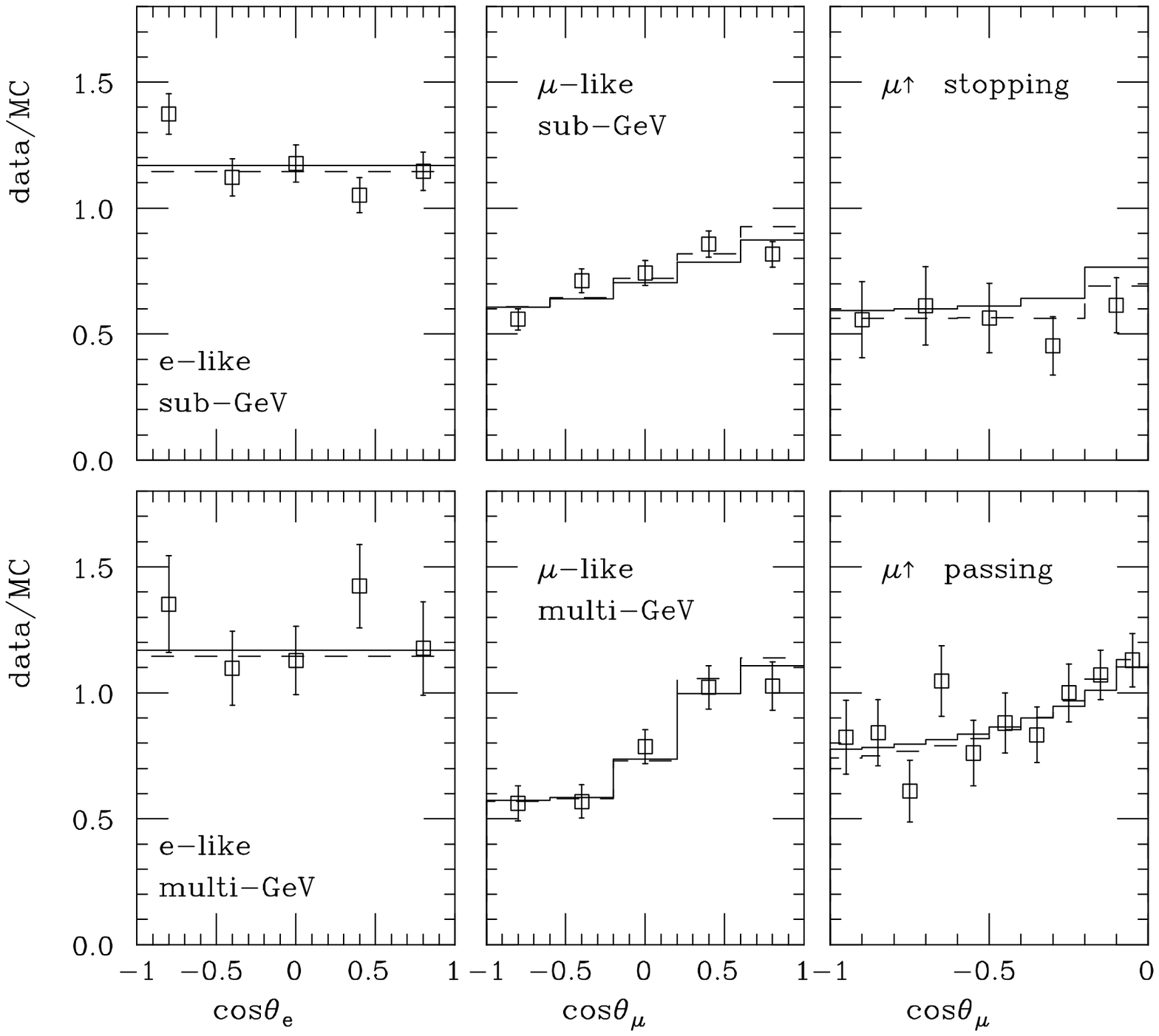}
\vspace{-3ex}

\caption[]{\label{fig:nudecay}
Comparison of neutrino oscillation (dashed lines) and neutrino decay (solid lines) model fits to SuperK $e$ and $\mu$ events from atmospheric neutrinos (from the first paper in Ref.~\cite{nu-decay}).}

\bigskip

\centering\leavevmode
\epsfxsize=4in\epsffile{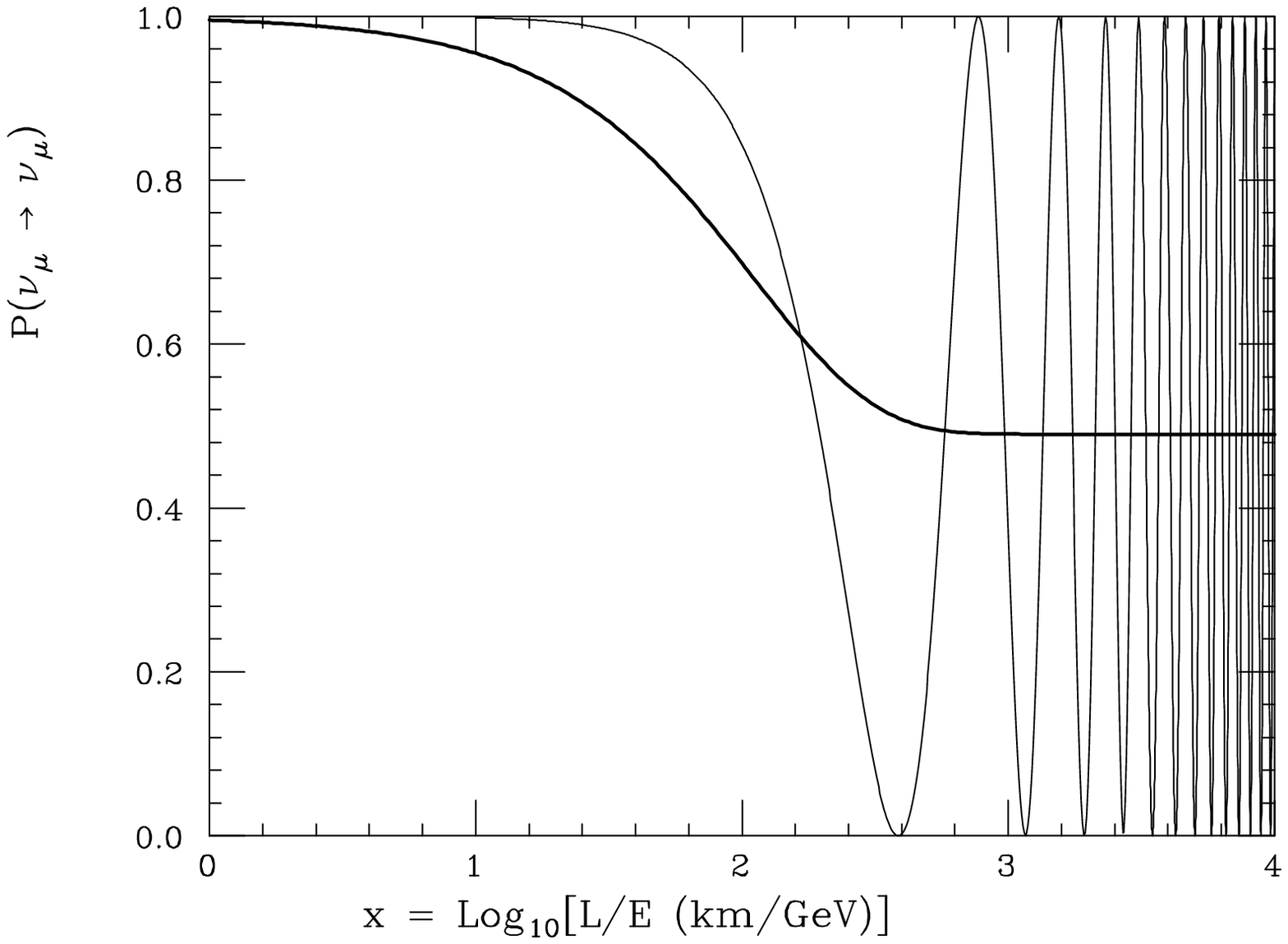}
\medskip

\caption[]{\label{fig:nudecay2}
Comparison of the $\nu_\mu\to \nu_\mu$ probabilities versus neutrino energy for the neutrino oscillation and decay models (from the first paper in Ref.~\cite{nu-decay}).}
\end{figure}

The Kamiokande, SuperKamiokande (SuperK), Macro and Soudan atmospheric neutrino measurements\cite{atmos1,atmos2} show a $\mu/e$ ratio that is about 0.6 of expectations. The SuperK experiment\cite{atmos1} has established the dependence of the $e$ and $\mu$ event rates on zenith angle, or equivalently the baseline $L$ (see Fig.~\ref{fig:depend}). $N(e)$ is independent of $L$ and validates the $\nu_e$ flux calculation (within 20\%). $N(\mu)$ depletion increases with $L$. The muon event distributions are will described by $\nu_\mu\to\nu_\tau$ vacuum oscillations with a $\nu_\mu$ {\em survival} probability
\begin{equation}
P(\nu_\mu\to\nu_\mu) = 1 - A_{\mu\tau} \sin^2 (1.27\delta m_a^2 L/E_\nu)
\end{equation}
with $\delta m_a^2 = 3.5\times 10^{-3}\rm\,eV^2$ and maximal or near maximal amplitude
\begin{equation}
A_{\mu\tau} = 1^{+0.0}_{-0.2}\quad (i.e.,\ |\theta_{32} - 45^\circ| < 13^\circ)
\,.
\end{equation}
With further data accumulation\cite{kajita} a slightly lower central value is now indicated ($\delta m_a^2 = 2.8 \times 10^{-3}\rm\,eV^2$). The $\nu_\mu\to\nu_\tau$ oscillations are not resolved due to smearing of $L$ and inferred $E_\nu$ values, and equally good fits to the SuperK data are found with oscillation and neutrino decay ($\nu_2\to\bar\nu_4 + J$) models\cite{nu-decay}, as shown in Fig.~\ref{fig:nudecay}. A comparison of the unsmeared $\nu_\mu\to\nu_\mu$ probability versus neutrino energy for the oscillation and decay models is shown in Fig.~\ref{fig:nudecay2}. For now we assume the simplest interpretation of the SuperK data, namely oscillations. We note that in the CHOOZ reactor experiment $\bar\nu_e$ disappearance is not observed at the $\delta m_a^2$ scale and the corresponding constraint on 3-neutrino mixing for $\delta m_a^2 = 3.5\times10^{-3}\rm\,eV^2$ is
\begin{equation}
A_{\mu e} < 0.2\,, \quad |U_{e3}| < 0.23\,, \quad \theta_{13} < 13^\circ \,.
\end{equation}
%

\section{Long-Baseline Experiments}

Long-baseline experiments are needed\cite{nu-decay,bernstein,BGRW,shrock,freund,rigolin,camp,CPV,derujula,moreCPV} to 
\def\theenumi{\roman{enumi}}
\def\labelenumi{(\theenumi)}

\begin{enumerate}
\addtolength{\itemsep}{-2.5mm}
\item confirm the atmospheric evidence for $P(\nu_\mu\to\nu_\mu)$ at accelerators;

\item resolve the leading $\nu_\mu\to\nu_\mu$ oscillation and exclude the neutrino decay possibility;

\item precisely measure $|\delta m_a^2|$;

\item exclude $\nu_\mu\to\nu_s$ disappearance, although SuperK has now shown that oscillations to sterile neutrinos are excluded at 99\% CL\cite{kajita};

\item measure $|U_{e3}|$ from $\nu_e\to\nu_\mu$ appearance, which requires a muon decay source for the neutrino beam;

\item determine the sign of $\delta m_a^2$ from matter effects in the Earth's crust;

\item search for CP violation\cite{freund,rigolin,camp,CPV,derujula,moreCPV}.

\end{enumerate}

The first long-baseline experiments will measure the energy dependence of the produced muons and measure the neutral-current to charged-current ratio, to partially address the first four issues listed above. The K2K experiment from KEK to SuperK is in operation, with a baseline $L=250$~km and mean neutrino energy $\left< E_\nu \right> = 1.4$~GeV. The MINOS experiment from Fermilab to Soudan, with $L=732$~km and possible energies of $\left<E_\nu\right> = 3, 6, 12$~GeV will begin in 2002. A 10\% precision on $|\delta m_a^2|$ may ultimately be possible at MINOS\cite{thesis}. The ICANOE\cite{icanoe} and OPERA\cite{opera} long-baseline experiments from CERN to Gran Sasso with $L\simeq 743$~km have been approved.

\section{Neutrino Factories}

Muon storage rings could provide intense neutrino beams ($\sim 10^{19}\mbox{--}10^{21}$ per year) that would yield thousands of charged-current neutrino interactions in a reasonably sized detector (10--50~kt) anywhere on Earth\cite{nufactories}. These neutrino factories would have pure neutrino beams ($\nu_e,\bar\nu_\mu$ from stored $\mu^+$ and $\bar\nu_e, \nu_\mu$ from stored $\mu^-$) with 50\% $\nu_e$ or $\bar\nu_e$ components. Detection of wrong-sign muons (the muons with opposite sign to the charge current from the beam muon neutrino) would signal $\nu_e\to\nu_\mu$ or $\bar\nu_e\to\bar\nu_\mu$ appearance oscillations. We now discuss the capability of a neutrino factory with $2\times10^{20}$ muons a year and a 10~kt detector to resolve the issues raised in the preceding section.

\begin{figure}[h]
\vspace{-2ex}

\centering\leavevmode
\epsfxsize=3.3in
\epsffile{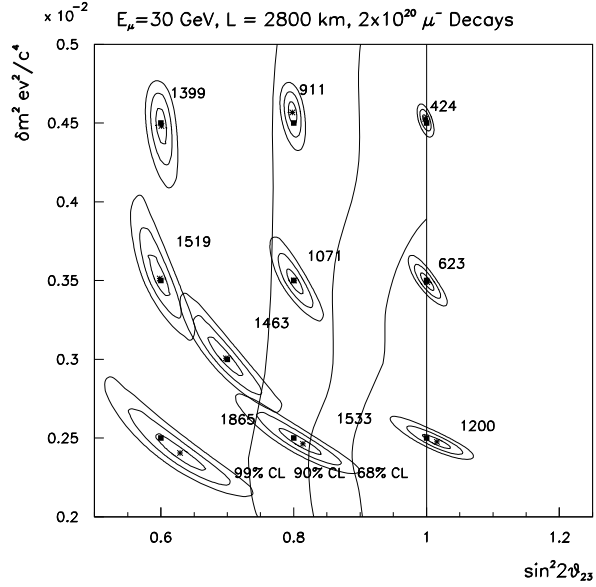}

\caption[]{\label{fig:precision}
Fit to muon neutrino survival distribution for $E_\mu = 30$~GeV and $L = 2800$~km for 10 pairs of $\sin^2 2\theta_{23}$ and $\delta m_{32}^2$ values; for each fit 1$\sigma$, 2$\sigma$ and 3$\sigma$ contours are shown. The generated points are the dark rectangles and the fitted values are the stars. The SuperK 68\%, 90\% and 95\% confidence levels are shown, and the predicted numbers of signal events for the points are given (from Ref.~\cite{BGRW}).}
\end{figure}

Figure \ref{fig:precision} shows the precision attainable in $\delta m_{32}^2$ and $\sin^22\theta_{23}$ parameters through $\nu_\mu$ survival measurements based on an $E_\mu=30$~GeV storage ring and a baseline of $L=2800$~km. A statistical precision of a few \% on $\sin^22\theta_{23}$ is possible in $\nu_\mu$ disappearance measurements. This accuracy in measuring $\sin^22\theta_{23}$ would differentiate the bimaximal model prediction\cite{bimax1,bimax2} of $\sin^22\theta_{23}=1$ from the democratic model prediction\cite{democ} of $\sin^22\theta_{23} = 8/9$.

\begin{figure}[b]
\vspace{-7ex}

\centering\leavevmode
\epsfxsize=2.8in\epsffile{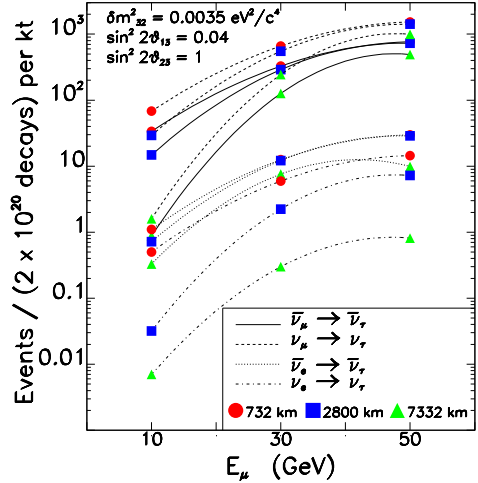}

\caption[]{\label{fig:taulepton}
Number of tau events from $\nu_\mu\to\nu_\tau$ oscillations in a 1~kt detector located at a distance $L=732$, 2800 and 7332~km for stored muon energy of $E_\mu = 10$, 30 and 50~GeV and a neutrino intensity of $2\times10^{20}$ (from Ref.~\cite{BGRW}).}
\end{figure}

Figure \ref{fig:taulepton} gives the tau-lepton yields at $E_\mu = 10$, 30 and 50~GeV for baselines of $L = 732$, 2800, and 7332~km in a 1~kt detector for an intensity of $2\times10^{20}$ neutrinos. There would be hundreds of events per year from $\nu_\mu\to\nu_\tau$ oscillations; however, the signal of $\nu_e\to\nu_\tau$ would be difficult with the $2\times10^{20}$ luminosity.

\begin{figure}[p]

\centering\leavevmode
\epsfxsize=2.6in\epsffile{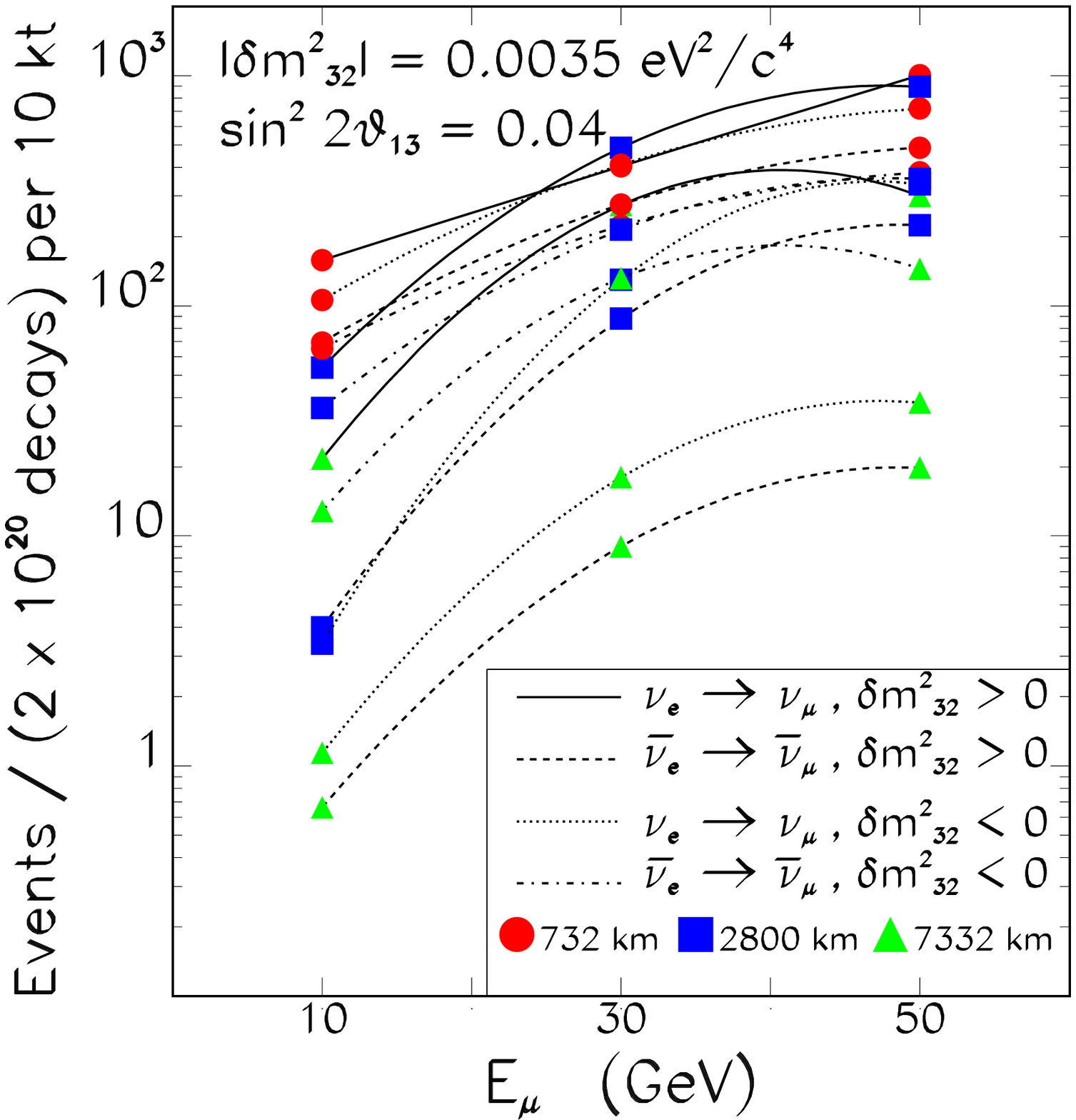}

\caption[]{\label{fig:wrongsign}
Wrong-sign muon event rates in a 10~kt detector at $L=2900$~km versus the stored muon energy $E_\mu$ (from Ref.~\cite{BGRW}).}

\bigskip

\centering\leavevmode
\epsfxsize=4.1in\epsffile{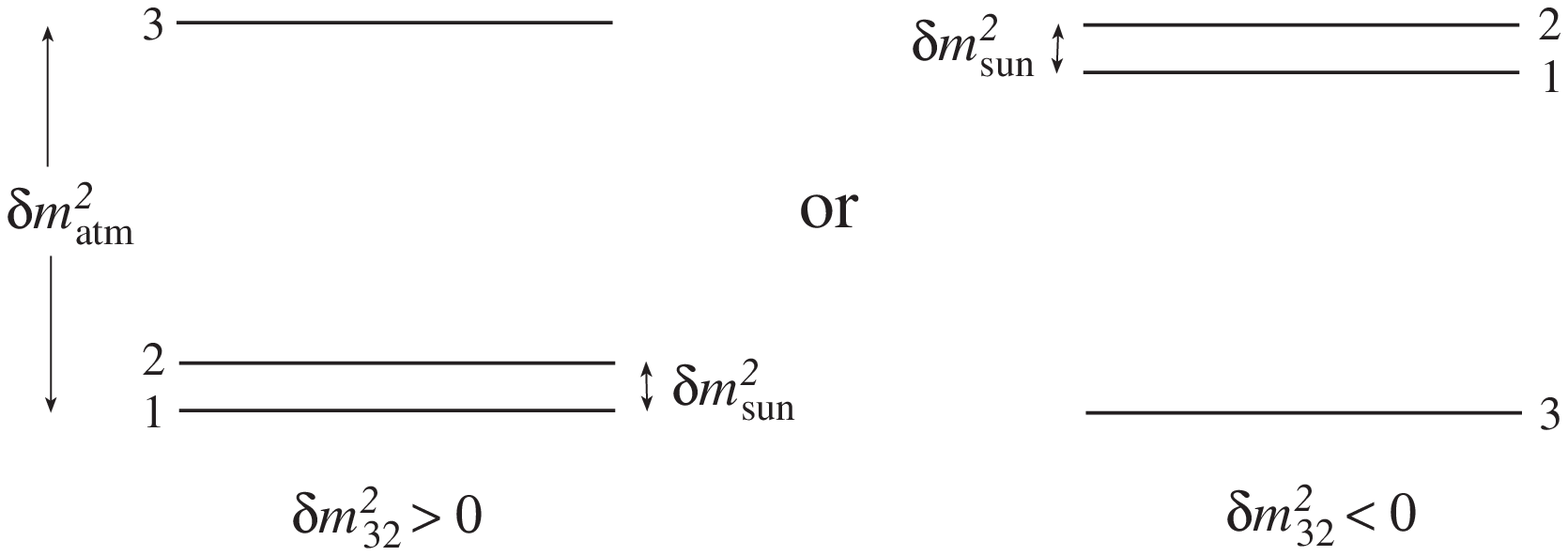}

\caption[]{\label{fig:masses}
Two possible patterns of three neutrino masses that can explain the atmospheric and solar anomalies.}

\bigskip

\centering\leavevmode
\epsfxsize=4.8in\epsffile{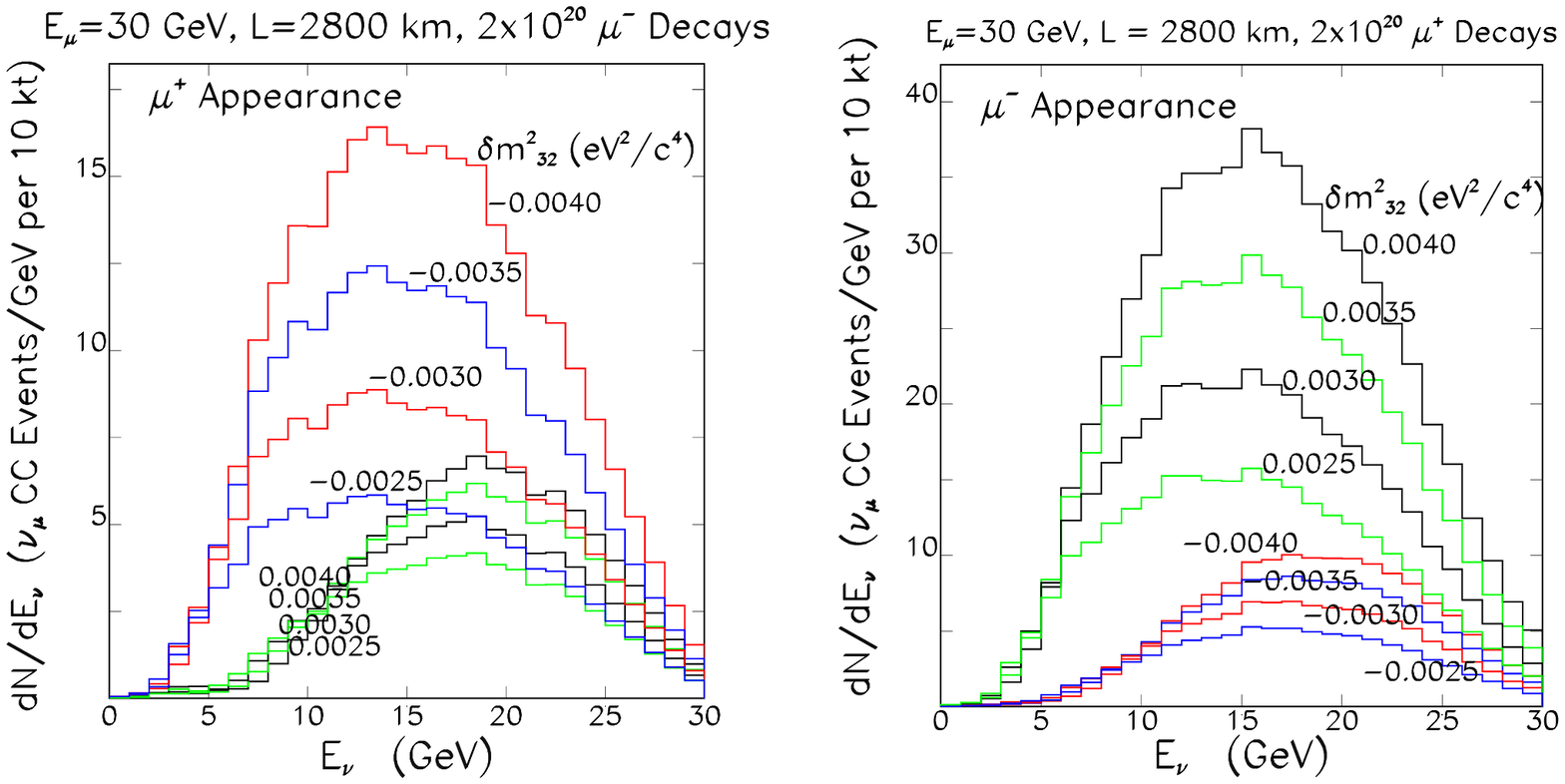}

\caption[]{\label{fig:proof}
Predicted energy distributions in CC events tagged by a wrong-sign (a)~positive muon from $\bar\nu_e\to\bar\nu_\mu$ oscillations, and (b)~negative muon from $\nu_e\to\nu_\mu$ oscillations, for various $\delta m^2$. The predictions are based on $2\times10^{20}$ decays, $E_\mu = 30$~GeV, $L=2800$~km, and a 10~kt detector (from Ref.~\cite{BGRW}).}
\end{figure}

The wrong-sign muon event rates are approximately proportional to $\sin^22\theta_{13}$. Figure~\ref{fig:wrongsign} shows the predicted numbers of events\cite{BGRW} with $\sin^22\theta_{13} = 0.04$ for a 10~kt detector at $L=2900$~km with an intensity of $2\times10^{20}$ neutrinos.

The observation of $\nu_e\to\nu_\mu$ and $\bar\nu_e\to\bar\nu_\mu$ appearance oscillations at baselines long enough to have significant matter effects will allow a determination of the sign of $\delta m_{32}^2$, and thus determine the pattern of the masses; see Fig.~\ref{fig:masses}. A proof of the principle that the sign of $\delta m^2$ can be so determined\cite{BGRW} is given in Fig.~\ref{fig:proof} for a baseline $L = 2800$~km. In $\mu^+$ appearance, $\delta m_{32}^2 >0$ gives a smaller rate and harder spectrum than $\delta m_{32}^2 < 0$, while the results are opposite in $\mu^-$ appearance.

\begin{figure}[t]
\centering\leavevmode
\epsfxsize=2.7in\epsffile{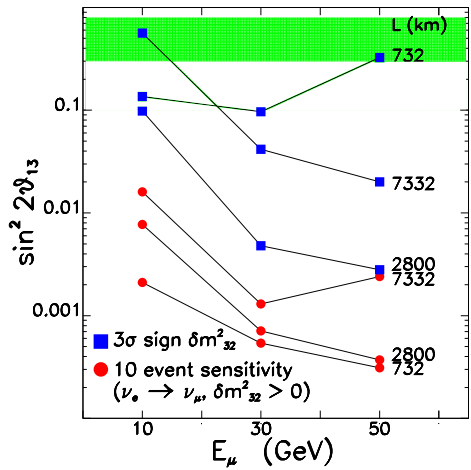}

\caption[]{\label{fig:sensitivity}
The value of $\sin^2 2\theta_{13}$ that yields, in a 10~kt detector
at $L = 2800$~km, (a) 10 events per $2 \times 10^{20} \mu^+$ decays (boxes), and
(b) a three standard deviation determination of the sign of $\delta m^2_{32}$
when the wrong-sign muon event rates for $2 \times 10^{20} \mu^+$ decays are
compared with the corresponding rates for $2 \times 10^{20} \mu^-$decays
(circles). The $\sin^2 2\theta_{13}$ sensitivity is shown versus $E_\mu$ and
$L$, as labelled (from Ref.~\cite{BGRW}).}
\end{figure}

In optimizing $E_\mu$ and $L$ for long-baseline experiments to find the sign of $\delta m_{32}^2$, $L = 732$~km is too short (matter effects are small) and $L = 7332$~km is too far (event rates are low)\cite{BGRW,rigolin}. The sensitivity to determine the sign of $\delta m_{32}^2$ improves linearly with $E_\mu$. There is a tradeoff between energy, detector size and muon beam intensity\cite{BGRW}. The $\nu_e\to\nu_\mu$ sensitivity on $\sin^22\theta_{23}$ and sign $\delta m_{32}^2$ for a 10~kt detector is shown\cite{BGRW} in Fig.~\ref{fig:sensitivity} for $E_\mu = 10$, 30, and 50~GeV.

\section{Solar Neutrino Oscillations}

The solar neutrino experiments\cite{solarexpts} sample different $\nu_e$ energy ranges and find different flux deficits compared to the Standard Solar Model (SSM)\cite{bahcall-sm} as follows:
\begin{equation}
\def\arraystretch{1.5}
\arraycolsep=.5em
\begin{array}{lll}
\nu_e {}^{71}{\rm Ga} \to {}^{71}{\rm Ge}\,e& \begin{array}{l}\rm GALLEX\\[-1ex] \rm SAGE\end{array} & \begin{array}{l}0.60\pm 0.06\\[-1ex] 0.52\pm 0.06\end{array}\\
\nu_e {}^{37}{\rm Cl}\to {}^{37}{\rm Ar}\, e&\rm\ Homestake& \ 0.33\pm 0.03\\
\nu e\to \nu e& \rm\ SuperK& \ 0.47\pm 0.02 \end{array}
\end{equation}
Thus the $\nu_e$ survival probability is inferred to be energy dependent.

Global oscillation fits\cite{solardata} have been made using floating $^8$B and hep flux normalizations which are somewhat uncertain in the SSM. The relative normalizations from the fits range from 0.5 to 1.2 for the $^8$B flux and 1 to 25 for the hep flux. These global fits include the data on
\begin{enumerate}
\addtolength{\itemsep}{-2mm}
\item total rates, assuming all the experiments are okay; the different Cl suppression ratio plays a vital role;

\item the night-day asymmetry, which is observed at the 2$\sigma$ level\cite{kajita}; the large angle matter solution gives night rates $>$ day rates;

\item seasonal dependence beyond $1/r^2$, which can occur for vacuum solutions.

\end{enumerate}
The oscillation analyses\cite{solardata} generally agree on the allowed $\delta m_{21}^2$ and $\sin^22\theta_{12}$ regions for acceptable solutions.
Typical candidate solar solutions are given in Table~\ref{tab:solar}. In the case of vacuum oscillations (VO) several discrete regions of $\delta m_{21}^2$ are possible.

\begin{table}[h]
\def\arraystretch{1.2}
\caption{\label{tab:solar}
Representative solutions to the solar neutrino anomaly.}
\begin{tabular}{lcc}
solution& $\sin^22\theta_{12}$& $\delta m^2_{21}\rm\ (eV^2)$\\
\hline
SAM& $\sim5\times10^{-3}$& $\sim5\times10^{-6}$\\
LAM& $\sim1$& $\sim3\times10^{-5}$\\
LOW& $\sim1$& $\sim10^{-7}$\\
VO& $\sim1$& $\sim10^{-10}$\\
\hline
\end{tabular}
\end{table}

\section{3-Neutrino Mixing Matrix}

Once the solar oscillation solution is pinned down, and $\theta_{12}$ is thus determined, we will have approximate knowledge of the mixing angles of the 3-neutrino matrix, with $\theta_{23} \sim \pi/4$ and $\theta_{13} \sim 0$ from the atmospheric and CHOOZ data. Upcoming experiments are expected to shed light on the solar solution. In the SNO experiment\cite{SNO}, which is now taking data, and the upcoming ICARUS experiment\cite{icarus}, the high energy $\nu_e$ CC events may distinguish LAM, SAM, and LOW solutions with large hep flux contributions from the VO or the SAM sterile neutrino solutions\cite{BKS99}. Also, the neutral-current to charged-current ratio will distinguish active from sterile oscillations. The Borexino\cite{borex}  experiment can measure the VO seasonal variation of the $^7$Be line flux\cite{7Be}. The KamLand reactor experiment\cite{kamland} to measure the $\bar\nu_e$ survival probability will be sensitive to the LAM and LOW solar solutions\cite{muray-kam}.

The CP phase $\delta$ may be measurable at a neutrino factory\cite{freund,rigolin,camp,derujula} if the solar solution is LAM. The CP violation comes in only at the sub-leading oscillation scale. The CP and T asymmetries for $\nu_\mu\leftrightarrow\nu_e$ oscillations are
\begin{eqnarray}
A^{CP}_{\alpha\beta} &=& { P(\nu_\alpha \rightarrow \nu_\beta) -
P(\bar\nu_\alpha \rightarrow \bar\nu_\beta) \over
P(\nu_\alpha \rightarrow \nu_\beta) +
P(\bar\nu_\alpha \rightarrow \bar\nu_\beta)} \,,\\
A^T_{\alpha\beta}
&=& { P(\nu_\alpha \rightarrow \nu_\beta) -
P(\nu_\beta \rightarrow \nu_\alpha) \over
P(\nu_\alpha \rightarrow \nu_\beta) +
P(\nu_\beta \rightarrow \nu_\alpha)} \,.
\end{eqnarray}
An apparent CP-odd asymmetry is induced by matter.

\section{Models}

For maximal mixing in both atmospheric and solar sectors, there is an unique mixing matrix\cite{bimax1} 
\begin{equation}
U = 
\left(\begin{array}{ccc}
1/\sqrt 2& -1/\sqrt 2& 0\\ 1/2& 1/2& -1/\sqrt 2\\ 1/2& 1/2& 1/\sqrt2
\end{array}\right) \,.
\end{equation}
In this bimaximal mixing model, there would be no CP-violating effects. However, because $U_{e3} = 0$, long-baseline experiments would have some sensitivity to the sub-leading LAM solar scale oscillations.

Many unification models\cite{RMreview,king} predict that the neutrino masses are Majorana and hierarchical, there is no cosmologically significant dark matter, and the SAM solar solution (small $\theta_{12}$ mixing) obtains. 

\section{Absolute Neutrino Masses}

Oscillation phenomena determine only mass-squared differences, leaving the absolute mass scale unknown. However, because the atmospheric and solar $\delta m^2$ values are $\ll (1\rm~eV)^2$, all mass eigenvalues are approximately degenerate if at the $\sim 1$~eV scale. Thus all neutrino mass eigenvalues are bounded\cite{BWW} by the tritium limit from the Troitsk and Mainz experiments,
\begin{equation}
m_j < 3{\rm\ eV\quad for}\ j = 1,2,3 \,.
\end{equation}
%

\begin{figure}[h]
\centering\leavevmode
\epsfxsize=.8in\epsffile{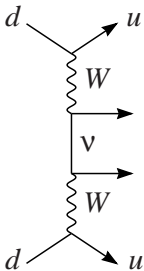}

\caption[]{\label{fig:0nbb}
Diagram for neutrinoless double beta decay via Majorana neutrino exchange.}
\end{figure}

Neutrinoless double-beta decay (0$\nu\beta\beta$) provides a probe of Majorana neutrino mass\cite{0nbb} via the diagram in Fig.~\ref{fig:0nbb}. The rate is proportional to the $\nu_e\nu_e$ element of the neutrino mass matrix. The present limit from the Heidelberg experiment\cite{baudis} is 
\begin{equation}
M_{\nu_e\nu_e} < 0.2{\rm\ eV} \times f\,,
\end{equation}
where the factor $f$ represents the theoretical uncertainty in the nuclear matrix elements, which might be as large as a factor of~3. The $0\nu\beta\beta$ limit translates to a bound on the summed neutrino Majorana masses of\cite{BW99}
\begin{equation}
\sum m_\nu < 0.75{\rm\ eV}\times f
\end{equation}
in the SAM solar solution. No similar constraints apply to the LAM, LOW or VO solutions where the bound can be satisfied by having opposite CP parity of $\nu_1$ and $\nu_2$ mass eigenstates. Future sensitivity down to
\begin{equation}
|M_{\nu_e\nu_e}| = 0.01\rm\ eV
\end{equation}
is expected\cite{future}, which would provide sensitivity down to
\begin{equation}
\sum m_\nu = 0.08{\rm\ eV} \times f
\end{equation}
in the SAM solution.

\section{Beyond 3 Neutrinos}

\begin{figure}[h]
\centering\leavevmode
\epsfxsize=2.3in\epsffile{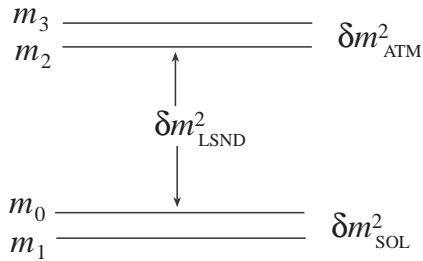}

\caption[]{\label{fig:4nu-mass}
Required mass pattern of two separated pairs to account for the LSND, atmospheric, and solar data. }
\end{figure}

The LSND evidence for $\nu_\mu\to\nu_e$ oscillations with $\delta m^2 \sim 1\rm~eV^2$, $\sin^22\theta\sim 10^{-2}$ requires a $\delta m^2$ scale distinct from the atmospheric and solar oscillation scales, and thus a sterile neutrino state would be needed to explain all the oscillation phenomena. Then to also satisfy limits from CDHS accelerator\cite{cdhs} and Bugey reactor\cite{bugey} experiments, the mass hierarchy must be two separated pairs\cite{doublets}, as shown in Fig.~\ref{fig:4nu-mass}. Such a scenario would allow even more interesting effects at a neutrino factory, such as large CP violation\cite{hattori}, since both the leading and sub-leading oscillation scales would be accessible. The MiniBooNE experiment\cite{miniboone} will settle whether the LSND evidence is real. Other interest in sterile neutrinos comes from $r$-process nucleosynthesis if it occurs in supernovae\cite{baha}.

\section{Neutrino Mass from Cosmology}

\begin{figure}
\centering\leavevmode
\epsfxsize=3.3in\epsffile{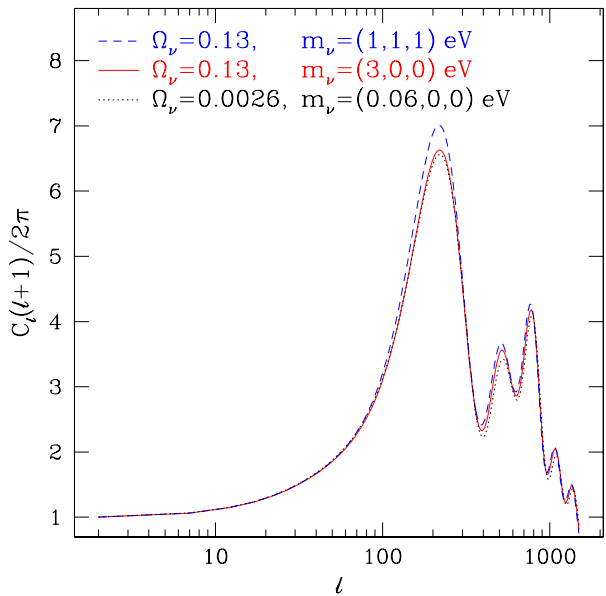}

\caption[]{\label{fig:cosmo}
Predicted power spectrum from three different mass scenarios.}
\end{figure}

Measurements of the power spectrum by the MAP and PLANCK satellites may determine $\sum m_\nu$ down to $\sim 0.4$~eV\cite{cosmo}. The heights of the acoustic peaks can also decide how the mass is distributed among the neutrino flavors\cite{falk}, as illustrated in Fig.~\ref{fig:cosmo}.

\section{Summary}

We have entered an exciting new era in the study of neutrino masses and mixing. From the SuperK evidence on atmospheric neutrino oscillations, we already have a surprising amount of information about the neutrino mixing matrix (near maximal $\sin^22\theta_{23}$ and near minimal $\sin^22\theta_{13}$). The SuperK, SNO, Borexino, KamLand, and ICARUS experiments are expected to differentiate among the candidate solar oscillation possibilities and determine $\sin^22\theta_{12}$. MiniBooNE will tell us whether a sterile neutrino is mandated. Neutrino factories will study the leading oscillations, determine the sign of $\delta m_a^2$, measure $U_{e3}$, and possibly detect CP violation. The GENIUS $0\nu\beta\beta$ experiment and the MAP and PLANCK satellite measurements of the power spectrum will probe the absolute scale of neutrino masses. There is a synergy of particle, physics, nuclear physics, and cosmology occurring in establishing the fundamental properties of neutrinos. A theoretical synthesis should emerge from these experimental pillars.

\section*{Acknowledgments}

I thank my collaborators K.~Whisnant, T.~Weiler, S.~Pakvasa, S.~Geer, R.~Raja, J.~Learned, P.~Lipari, and M.~Lusignoli. This research was supported in part by the U.S. Department of Energy under
Grant No.~DE-FG02-95ER40896  and in part by the
University of Wisconsin Research Committee with funds granted by the Wisconsin
 Alumni Research Foundation.

\end{document}